Enhancement of critical current density in (Ba,Na)Fe$_2$As$_2$ round wires using high-pressure sintering

Sunseng Pyon[1], Daisuke Miyawaki[1], Tsuyoshi Tamegai[1], Satoshi Awaji[2], Hijiri Kito[3], Shigeyuki Ishida[3], Yoshiyuki Yoshida[3]

[1]Department of Applied Physics, The University of Tokyo, Bunkyo, Tokyo 113-8656, Japan
[2]High Field Laboratory for Superconducting Materials, Institute for Materials Research, Tohoku University, Sendai 980-8577, Japan
[3]Electronics and Photonics Research Institute, National Institute of Advanced Industrial Science and Technology, Tsukuba, Ibaraki 305-8568, Japan

**Abstract**

We fabricated (Ba,Na)Fe$_2$As$_2$ superconducting wires using the powder-in-tube method and hot isostatic pressing. By improving powder synthesis processes compared with previous studies, highly pure raw materials for the wire fabrication were obtained. The largest transport critical current density ($J_c$) reached 40 kAcm$^{-2}$ at $T$ = 4.2 K under a magnetic field of 100 kOe. This value exceeds not only the value of transport $J_c$ of the previous (Ba,Na)Fe$_2$As$_2$ wire but also those of all iron-based superconducting round wires. Improvements of polycrystalline powder synthesis play a key role for the enhancement of $J_c$. Furthermore, it was clarified that higher densification by high-pressure sintering and the texturing of grains in the core of the wire due to drawing also increased $J_c$ effectively. Advantages of (Ba,Na)Fe$_2$As$_2$ wires compared with (Ba,K)Fe$_2$As$_2$ wires are also discussed.

Keywords: iron-based superconducting round wires, critical current density, powder-in-tube, HIP, (Ba,Na)Fe$_2$As$_2$

**Introduction**

In the past decades, iron-based superconductors (IBSs) and their wires and tapes have been studied as attractive candidates for future high-field applications of superconductors as described in several review articles [1-8]. They exhibit high critical temperature, $T_c$, large upper critical field, $H_{c2}$, and relatively low anisotropy, $\gamma$, compared with cuprate superconductors, which are advantageous for good high-field performance. Among IBSs, 122-type compounds of $(AE',K)Fe_2As_2$ ($AE'$ = Ba, Sr) [9,10] are considered to be promising materials for practical use with their high $T_c$ (~38 K) [9,10], high $H_{c2}$ (> 650 kOe) [10-13], and small $\gamma < 2$ [11-13]. These characteristics as well as excellent grain-boundary nature make them attractive compared with other candidates for future applications such as $MgB_2$ or cuprates [6]. In addition, very large critical current density, $J_c$, over 1 $MAcm^{-2}$ below 5 K at self-field was achieved in single crystals of $(AE',K)Fe_2As_2$, and it can be enhanced further by introducing artificial defects [14-18]. Using these promising materials, wires and tapes have been fabricated by powder-in-tube (PIT) method. Some studies on IBS wires suggest that the low transport $J_c$ and its strong magnetic field dependence are caused by weak links between superconducting grains. In recent studies, pressing techniques such as hot press for textured tapes [19-24] or hot isostatic press (HIP) treatment for round wires [17, 25-32] increased the transport $J_c$ significantly. The transport $J_c$ at 4.2 K under a high field of 100 kOe for tapes and wires has reached 150 and 38 $kAcm^{-2}$, respectively [24,31,32]. Not only the enhancement of $J_c$ but also some remarkable progress relevant to practical applications, such as the fabrication of long wires [33], multifilamentary fabrication [21,34,35], and superconducting joints for $(AE',K)Fe_2As_2$ tapes [36], have been reported.

In contrast to K-doped 122-type materials, the investigation of $J_c$ of single crystals and the fabrication of wires or tapes of Na-doped 122-type material, $(AE,Na)Fe_2As_2$ ($AE$ = Ba, Sr, or Ca) [37-40], have not been performed extensively, in spite of their comparable $T_c$ (~34 K) to K-doped ones. Only recently, some studies of tapes and wires of $(AE,Na)Fe_2As_2$ have been reported [41-47]. Iyo *et al*. reported that the transport $J_c$ for $(Sr,Na)Fe_2As_2$ tapes at 20 K under a magnetic field of 25 kOe was superior to that for $(AE',K)Fe_2As_2$ tapes [41]. The transport $J_c$ for $(Sr,Na)Fe_2As_2$ tapes of 26 $kAcm^{-2}$ at 4.2 K under 100 kOe, and that for $(Ba,Na)Fe_2As_2$ tapes of 40 $kAcm^{-2}$ at 4.2 K under 40 kOe, were also reported [42,45]. Very recently, $J_c$ in $(Ba,Na)Fe_2As_2$ HIP round wire reached 150 and 20 $kAcm^{-2}$ at self-field and 100 kOe, respectively [46-48]. These values are more than fifty percent of the largest $J_c$ in $(Ba,K)Fe_2As_2$ round wires, demonstrating the excellent performance of $(Ba,Na)Fe_2As_2$ wires.

For further enhancement of $J_c$ in $(Ba,Na)Fe_2As_2$ wires or tapes, weak links between superconducting grains caused by impurity phase should be removed. Compared with K-doped materials, synthesis conditions of polycrystalline powders of $(Ba,Na)Fe_2As_2$ and the sintering conditions of wires had not been optimized yet. It is reported that compositional inhomogeneities of materials and the presence of impurity phases in the core of wire decrease $J_c$ significantly, not only for 122-type compounds but also for other IBS materials such as 1111-type [49], 11-type [50]. 1144-type [51-53], and 21113–type compounds [54,55]. For $(Ba,Na)Fe_2As_2$ compounds, it is reported that the quality of polycrystalline samples is strongly affected by the sintering temperature around 770-850$^{\circ}$C [47]. Optimizations of powder-synthesis conditions had been demanded.

In this paper, we focus on the synthesis of high-quality $(Ba,Na)Fe_2As_2$ powders and the fabrication of round wires using them. Round shapes of wires are more suitable than textured flat tapes for a wide applications. So round wires processed by the HIP technique are promising for high-field applications. By using a higher quality polycrystalline powder of $(Ba,Na)Fe_2As_2$, superconducting wires were fabricated by PIT and HIP method. Obtained HIP wires have been thoroughly characterized and both the transport and magnetic $J_c$ are measured. The largest transport $J_c$ reached 40 kAcm$^{-2}$ at $T$ = 4.2 K under a magnetic field of 100 kOe. This value exceeds not only the value of transport $J_c$ of the previous $(Ba,Na)Fe_2As_2$ wire but also those of all iron-based superconducting round wires. Details of the powder synthesis process, wire fabrications, and characterizations of fabricated wires are shown and then discussed.

**Experimental methods**

Superconducting wires of $(Ba,Na)Fe_2As_2$ were fabricated by *ex situ* powder-in-tube (PIT) method. In the present studies, polycrystalline powders of $(Ba,Na)Fe_2As_2$ were prepared by two different ways of solid-state reaction. In the first method, Ba pieces, Na ingots, Fe powder, and As pieces were used as starting materials. The starting mixture contained 15% excess Na and 5% excess As, to compensate the loss of elements. After filing Ba pieces from the chunk and crushing of As pieces, starting materials were mixed for more than 10 h in Ar atmosphere using a planetary ball-milling machine and densely packed into a niobium tube. The niobium tube was then put into a stainless steel tube and sealed in argon-filled glove box for heat treatment for 30 h at 800, 850, and 900°C. It was then ground into powder using an agate mortar in an argon-filled glove box. We designate the powder that was heated at 850°C as the #1 powder. In the second method, BaAs, NaAs and Fe$_2$As were used as precursors to synthesize $(Ba,Na)Fe_2As_2$. For BaAs, a flat-plate Ba was finely cut into small pieces. They were weighed out with As in total of ~2 g, and sealed in an evacuated quartz tube. It was heated for 20 h at 700°C, after intermediate heating for 10 h at 500°C. For NaAs, Na and As were weighed out in total of ~1 g, and they were put in an alumina crucible, and then put in a stainless steel tube, and sealed in an argon-filled glove box. After intermediate heating for 10 h at 500°C, it was heated for 24 h at 800°C. For Fe$_2$As, Fe and As were weighed out in total of ~3 g, and sealed in an evacuated quartz tube. After an intermediate heating for 10 h at 500°C, it was heated for 40 h at 800°C. The prepared precursors were weighed to a composition of BaAs:NaAs:Fe$_2$As = 0.6:(0.4+α):1, thoroughly mixed. Here, α( = 0.1) is additional NaAs content to compensate the loss of Na and As during the synthesis. Polycrystalline powders were synthesized for 24 h at 820°C in an alumina crucible which was enclosed in an argon-filled stainless steel tube. We designate thus obtained powder as the #2. The obtained two kinds of powders were ground and filled into silver tubes with outer and inner diameters of 4.5 mm and 3 mm, respectively. Ag tube filled with the #1 powder was cold-drawn using dies with circular holes. On the other hand, the tube filled with the #2 powder was swaged using a rotary swaging machine. These wires were formed into a round shape with a diameter of ~1.5 mm for the #1 wire, and ~1.35 mm for the #2 wire, respectively. After cutting them into short pieces, one of the pieces was put into 1/8 inch copper tube and redrawn into a square shape with a groove roller down to a diagonal dimension of 1.2 mm. After the drawing process, both ends of the wire were sealed using an arc welder. The sealed wires were sintered using the HIP technique. Wires were heated for 4 h at 700°C in an argon atmosphere under different pressures of 0.1–175 MPa. HIP processes under the highest pressure 175 MPa were performed at National Institute of Advanced Industrial Science and Technology or Metal Technology Co. Ltd.. Others were performed at the University of Tokyo. Obtained HIP wires using the #1 and #2 powders are designated as the #1 and #2 wires, respectively. In order to evaluate the transport $J_c$, a DC electric current up to 100 A was delivered through the wire at a ramping rate of 50–100 $Amin^{-1}$. In order to minimize the effect of Joule heating at the current leads, measurements were performed in liquid helium. The current density when the electric field exceeds 1 $\mu Vcm^{-1}$ is defined as the transport $J_c$. To evaluate $J_c$, critical current $I_c$ was divided by the area of superconducting core of the HIP wire. The $I_c$ measurements in high magnetic fields up to 140 kOe were carried out by using the 15T-SM at the High Field Laboratory for Superconducting Materials, IMR, Tohoku University. Current–voltage ($I$–$V$) characteristics were measured by the four-probe method with solder for contacts. The bulk magnetization of a short piece of the wire was measured to characterize $T_c$ and magnetic $J_c$ by a superconducting quantum interference device magnetometer (SQUID, MPMS-5XL, Quantum Design). Vickers hardness, $Hv$, was measured on the polished surface of the wire core. For MO imaging, the HIP wire was cut and the transverse cross section was polished with lapping films. An iron–garnet indicator film was placed in direct contact with the sample and the whole assembly was attached to the cold finger of a He-flow cryostat (Microstat-HR, Oxford Instruments). MO images were acquired by using a cooled-CCD camera with 12-bit resolution (ORCA-ER, Hamamatsu). Powder X-ray diffraction (XRD) with Cu-K$\alpha$ radiation (Smartlab, Rigaku) for the polycrystalline powders and the core of the HIP wires were carried out for the phase identification and evaluation of texturing of the core of the HIP wire. The wire core was observed using scanning electron microscope (S-4300, Hitachi High Technologies), and elemental mappings were conducted using energy-dispersive X-ray spectroscopy (EDX) with EMAX x-act (HORIBA).

**Results and discussions**

Polycrystalline powders for superconducting wires were characterized by magnetization and XRD measurements. Onsets of diamagnetization of both the #1 and #2 powders were detected at approximately 34.5 K as shown in Fig. 1(a). Although the drop of the magnetization of the #1 powder is slightly sharper than that of the #2 powder, reasonably sharp transition indicates that the quality of both powders are higher than the previous study [46-48]. XRD patterns of the #1 and #2 powders shown in Fig. 1(b) have strong peaks of $(Ba,Na)Fe_2As_2$ phase and do not show peaks of impurities such as FeAs or $Fe_2As$. It is noteworthy that powders mixed by ball milling method and sintered at 800 or 900°C show also sharp magnetization transitions and clear XRD patterns of $(Ba,Na)Fe_2As_2$, although that sintered at 900°C shows a small amount of impurity peaks (data not shown).

Superconducting properties of HIP wires using the #1 and #2 powders were examined. Among more than five wires for the #1 wires and two wires for #2 wires that have been measured for their $J_c$, data showing the magnetic field dependence of maximum transport $J_c$ are summarized in Fig. 2(a). The most important achievement is the largest transport $J_c$ value among IBS superconducting HIP wires under high magnetic field of 100 kOe. As shown in Fig. 2(a), $J_c$ under a magnetic field of 100 kOe reach 40 and 34 kAcm$^{-2}$ for the #1 and #2 wires, respectively. It should be noted that the HIP wire using the powder mixed by ball milling method and sintered at 800°C also shows large transport $J_c$ ~ 38 kAcm$^{-2}$ under 100 kOe at 4.2 K (data not shown). These values are larger than the value of previous studies on $(Ba,Na)Fe_2As_2$ wires [46-48]. Furthermore, the transport $J_c$ of the #1 wire under 100 kOe slightly exceeds the previous largest value of transport $J_c$ (~ 38 kAcm$^{-2}$) among all round wires achieved in $(Ba,K)Fe_2As_2$ [32]. The self-field transport $J_c$ reach 95 or 119 kAcm$^{-2}$ for the #1 and #2 wires, respectively. It is noteworthy that the transport $J_c$ under a magnetic field of 1 kOe reach 127 and 140 kAcm$^{-2}$ for the #1 and #2 wires, respectively, which are larger than the $J_c$ under the self-field. Such behavior was also discussed by Hecher *et al.* and may be caused by grain size effect [29]. The magnetic $J_c$ of the same wires evaluated from the irreversible magnetization using the extended Bean model [14] is also plotted in Fig. 2(b). We used the formula of the extended Bean model $20\Delta M/a(1-a/3b)$, where $\Delta M$[emu/cm$^3$] is $M_{down}$ - $M_{up}$. $M_{up}$ and $M_{down}$ are the magnetization when sweeping the field up and down, respectively, and $a$[cm] and $b$[cm] are the lateral dimensions of the core, approximated by a rectangle ($a < b$), keeping the same area of the actual core. The magnetic $J_c$ under the self-field at 4.2 K in the #1 and #2 wires was 204 and 327 kAcm$^{-2}$, respectively. These magnetic $J_c$ values are more than twice larger than those of transport $J_c$ under the self-field. On the other hand, at higher magnetic fields, magnetic and transport $J_c$ are comparable. Under the field of 30 kOe, the magnetic $J_c$ of the #1 and #2 wires are 62 and 55 kAcm$^{-2}$, and the transport $J_c$ of them are 61 and 47 was kAcm$^{-2}$. Figure 2(c) shows the temperature dependence of magnetization for the HIP wires at $H$ = 5 Oe. Both magnetization curves show sharp transitions around $T_c$. $T_c$ in the #1 or #2 wires were 33.6 K and 33.0 K, respectively, which are slightly lower than those in powders (~34.5 K). The small reduction of $T_c$ after the wire fabrication may be caused by the degradation in the drawing process and incomplete recovery by the successive sintering, as observed in (Ba,K)Fe$_2$As$_2$ wire [28].

To discuss texturing in the core of the #1 and #2 wires, they were cut and polished, and different surfaces of the wire core were measured by XRD. Two planes of each wires were measured which were parallel (longitudinal), and perpendicular (transverse) to the current flow direction as shown in the insets of Figs. 3(a)-(d). Intensities for (002) and (103) XRD peaks for each surfaces are shown and compared in Figs. 3(a) and (b) for the #1 wire, in Figs. 3(c) and (d) for the #2 wire. It should be noted that the intensity in the left panel of Figs. 3(b) and (d) is multiplied by a factor of 2.2 to normalize the incident X-ray intensity for the comparison with the right panel of Figs. 3(b) and (d). This factor is evaluated from the diameter of collimated X-ray beams of 0.8 mm, and the major axis of the beam spot along the beam line on the wire surfaces is 6.9 mm ($2\theta = 13.4^\circ$) and 3.0 mm ($2\theta = 30.7^\circ$). Here, we define the ratio between the intensities of (002) peak and that of (003) peak, $r \equiv I(002)/I(103)$. The value of $r$ expresses the degree of texturing of grains in the wire core. It is known that in the case of tapes, where grains are well-textured along the $c$-axis, the $r$ becomes much larger than unity [19]. When the X-ray is reflected from the surface parallel to the current direction, $r$ is ~0.27 and ~0.20 for the #1 and #2 wires, respectively, as shown in Figs. 3(a) and (c). On the other hand, as shown in Figs. 3(b) and (d), when the X-ray is reflected from the surface perpendicular to the current, the intensity of the (002) peak is almost comparable to the noise level and the value of $r$ is less than ~0.02 for both the #1 and #2 wires. The negligible intensity of the (002) peak shown in Figs. 3(b) and (d) indicates that the fraction of grains with their $c$-axis perpendicular to the current is significantly low. Similar reduction of the intensity of peaks was observed in drawn wires of (Ba,K)Fe$_2$As$_2$ [32]. It is suggested that grains in the core are concentrically textured during the drawing or swaging, and groove-rolling processes.

Next, for detailed analyses of the impurity phases, XRD measurements of the wires were performed. The #1 and #2 wires were polished as shown in the insets of Figs. 3(a) and (c). Figure 4 shows the XRD patterns of the #1 and #2 wires, which are normalized by the intensity of the (103) peak of (Ba,Na)Fe$_2$As$_2$ at $2\theta \sim 30.7^\circ$. XRD pattern of the #1 powder, which is described in Fig. 1(b) is also shown as a reference of pure (Ba,Na)Fe$_2$As$_2$. XRD patterns of the #1 wire do not show impurity peaks except for the peaks of the Cu and Ag sheath, or peaks from Cu-K$\beta$ X-ray. This indicates that the amount of impurities in the core of the #1 wire is negligible. On the other hand, XRD pattern of the #2 wire shows a small but non-negligible amount of peaks of impurities of FeAs, Fe$_2$As, Ba(OH)$_2$, Na$_2$O$_2$(H$_2$O)$_8$, and unknown compounds. We interpret that Ba and Na rich areas as impurity phases reacted with moisture in air and generate Ba(OH)$_2$ and Na$_2$O$_2$(H$_2$O)$_8$ during XRD measurements. These results suggest a small amount of decomposition of (Ba,Na)Fe$_2$As$_2$ phase in the core of the #2 wire during the wire fabrication process. The presence of impurity phases in the core of the #1 and #2 wires were also evident from the observation of optical micrographs and SEM-EDX analyses. Figures 5 (a) and (b) show the optical micrographs of the core of the #1 and #2 wires. Optical micrographs of cross sections of the wire core are also shown in the insets of Figs. 5(a) and (b), respectively. Cu, Ag, and $(Ba,Na)Fe_2As_2$ core regions are identified, and no voids are observed. We find that there are bright regions in the core of both wires and the area of these bright regions in the core of the #2 wire is larger than that in the core of the #1 wire. Figures 5(c) and (d) show SEM images of the core of wires corresponding to blue squares in Figs. 5(a) and 5(b), respectively. Grains of 5~10 μm size observed in the SEM images correspond to the bright islands in the optical micrograph. From EDX analyses and compositional mappings of Ba, Na, Fe, and As, as shown in Figs. 5(e) and (f), it is clarified that the bright areas are mainly composed of impurities of FeAs or $Fe_2As$. In addition, in the core of the #2 wire, $Ba(OH)_2$ phase was also detected as shown in the compositional mapping of Ba in Fig. 5(e). These results are consistent with the results of XRD analyses shown in Fig. 4. Compositional analyses by EDX were also performed for $(Ba,Na)Fe_2As_2$ phase. The average Na content $x$ is approximately 0.38 both the #1 and #2 wire core, which is smaller than the nominal composition of $x \sim 0.46$-0.5 considering the excess Na for the #1 powder and NaAs for the #2 powder, respectively.

Compared with former studies on $(Ba,Na)Fe_2As_2$ wires [46-48], the quality of prepared polycrystalline powders for raw materials of wires and superconducting properties of HIP wires are significantly improved in the present study. There are several differences of powder synthesis process of the #2 powder compared with that of powder in the previous study [46,48], such as the increase in sintering temperature from 770°C to 820°C, the increase in the additional NaAs content α from 0.05 to 0.1, the reduction of a number of sintering from twice to once. The #1 powder and related powders, in which raw elements had been mixed by ball milling method, were also fired only once at a higher temperature of 800~900°C. Considering these differences of synthesis conditions and the fact that the analyzed Na content is less than the nominal content of $x$, sintering at above 800°C for a long enough time and the compensation of the loss of Na during the synthesis have played key roles for preparing $(Ba,Na)Fe_2As_2$ powders with negligible impurities. In spite of these purifications, however, a small amount of impurities were detected in the cores of both HIP wires, in particular in the #2 wire. Considering the fact that the final sintering conditions are the same for both the #1 and #2 wires, the difference of the distributions of impurities in the cores of the #1 and #2 wires may be attributed to the different quality of the #1 and #2 powders, which were not detected by XRD analyses. The presence of much more impurities in the core of the #2 wire compared with that of the #1 wire may be the main reason for the smaller transport $J_c$ under high magnetic fields above 10 kOe. Further improvements of the process of polycrystalline powders are demanded to increase $J_c$ in the wires. On the other hand, a small difference in $T_c$ and the degree of texturing between the #1 and #2 wire should also affect the transport $J_c$ in these wires, similar to case of the (Ba,K)Fe$_2$As$_2$ HIP wire [32]. There is also a room for the improvement of the process of wire fabrication for further increase in $J_c$. In addition, it is still not clear why both the transport $J_c$ and magnetic $J_c$ of the #2 wire under lower magnetic fields below 10 kOe were larger than those of the #1 wire. One of the possible scenarios is that impurities in the core of the #2 wire work as pinning centers rather than disadvantageous weak links between superconducting grains under magnetic fields below 10 kOe. If it is the case, a further increase in $J_c$ at low magnetic fields can be realized by intentional addition of impurities, which work as pinning centers. However, the pinning at low fields below 10 kOe cannot be explained only by relatively large impurity phases detected by optical and SEM images shown in figure 5. Extensive microstructural analyses including TEM and SEM observations on impurity phases and/or defects would be necessary to elucidate the pinning mechanism at low fields.

Next, using the #1 wire, we evaluated the effect of high-pressure sintering on physical properties in the HIP wires. The local $J_c$ distribution in the core of the #1 wire was evaluated from MO measurements. An optical micrograph of the transverse cross sections of the #1 wire is shown in Figure 6(a). Figures 6(b) and (c) show MO images of the transverse cross section of the core of the #1 wire in the remanent state at 5 and 30 K, respectively. Smooth variations of the trapped magnetic field in the #1 wire at not only 5 K but also 30 K indicate the presence of a uniform bulk current flowing in the wire core across grains in a wide temperature range below $T_c \sim 33.6$ K. This suggests that grains are well connected and they have homogeneous compositions. Figure 6(d) shows the magnetic induction profiles along the red line in Fig. 6(b). As the temperature is increased towards the $T_c$, the intergranular critical current in the #1 wire decreases only gradually. Such behavior also indicates that a uniform bulk current flows in a broad temperature range below the $T_c$. These observations suggest that the quality of the core of (Ba,Na)Fe$_2$As$_2$ is high and homogeneous, and that weak links between grains are improved by sintering using the HIP technique.

Effects of sintering pressure on physical properties in the HIP wire were also evaluated. Figure 7(a) shows the temperature dependence of the magnetization of the HIP wires using the #1 powder sintered at 700$^{\circ}$C for 4 h at pressures of 0.1, 9, and 175 MPa. Although the onset $T_c$ is almost the same, it is clear that as the sintering pressure increases, the superconducting transition becomes sharper. It may be caused by the difference in the degree of shielding current flow due to different densification of the core. Other possible reason is related to the degradation of (Ba,Na)Fe$_2$As$_2$ in the wire core during the fabrication process and partial recovery of superconducting properties during the sintering, reported in the case of (Ba,K)Fe$_2$As$_2$ wires [28]. A higher sintering pressure should be more effective for the recovery. The magnetic $J_c$ also shows clear sintering pressure dependence. Figure 7(b) shows the field dependence of the magnetic $J_c$ in these wires. The magnetic $J_c$ under the self-field at 4.2 K in wires sintered at 175, 9, and 0.1 MPa were 204, 89, and 71 kAcm$^{-2}$, respectively. The magnetic $J_c$ under the field of 40 kOe at 4.2 K of those wires were 52, 15, and 8.5 kAcm$^{-2}$, respectively. These values are summarized in Fig. 7(c) as sintering pressure dependence of the magnetic $J_c$. $H$v for these wires were also measured, as plotted in Fig. 7(d). The $H$v of these wires sintered at 175, 9, and 0.1 MPa were 216, 160, and 77, respectively. Both the magnetic $J_c$ and $H$v increased with the increase in the sintering pressure, which is similar to the case of (Ba,K)Fe$_2$As$_2$ and CaKFe$_4$As$_4$ HIP wires [32,51]. These results suggest that a high core density and the resulting suppression of weak links by high-pressure sintering increase transport $J_c$ effectively.

From MO analyses and a systematic investigation on the sintering-pressure dependence of physical properties of the (Ba,Na)Fe$_2$As$_2$ HIP wire, it is clarified that high-pressure sintering with suppression of weak links between the superconducting grains works effectively for the increase in $J_c$ in these wires. These tendency are similar to those in (*AE'*,K)Fe$_2$As$_2$ HIP wires. It has been demonstrated that (Ba,Na)Fe$_2$As$_2$ wires are very promising materials for future applications, as well as (*AE'*,K)Fe$_2$As$_2$ compounds. Furthermore, there are several distinct features of (*AE*,Na)Fe$_2$As$_2$ wires, as compared to those of (*AE'*,K)Fe$_2$As$_2$ wires. First, (*AE*,Na)Fe$_2$As$_2$ wires and tapes show weaker field dependence of the transport $J_c$ compared to that of (*AE'*,K)Fe$_2$As$_2$. The ratio of transport $J_c$ at a high field of 100 kOe to that at a low field of 1 kOe in the #1 wire is ~0.31, which is larger than that in the (Ba,K)Fe$_2$As$_2$ HIP wire (~0.19) [32]. Relatively small field dependence of $J_c$ is also demonstrated in the previous studies on (Ba,Na)Fe$_2$As$_2$ wires [46-48] and (Sr,Na)Fe$_2$As$_2$ tapes [42-44]. Obviously, this is a very advantageous property for high-field applications. Second, the $J_c$ of (*AE*,Na)Fe$_2$As$_2$ at high temperatures below 20 K is larger than that of (*AE'*,K)Fe$_2$As$_2$. Figure 8(a) shows the magnetic field dependence of the magnetic $J_c$ of the #1 wire at several temperatures. Magnetic $J_c$ decreases monotonically with increasing temperature at magnetic fields below 50 kOe. Temperature dependence of the magnetic $J_c$ of the #1 wire under the self-filed and at 40 kOe is plotted in Fig. 8(b). As a reference, the same sets of data on the (Ba,K)Fe$_2$As$_2$ HIP wire, which is the wire reported in ref. [32] were also measured and is plotted in Fig. 8(b). Although the magnetic $J_c$ of the #1 wire under the self-field is smaller than that of the (Ba,K)Fe$_2$As$_2$ HIP wire at all temperatures, it becomes larger than that of the (Ba,K)Fe$_2$As$_2$ at 40 kOe below 20K. This higher performance of the #1 wire compared with the (Ba,K)Fe$_2$As$_2$ HIP wire below 20 K is more clearly demonstrated in the inset of Fig. 8(b), where the normalized temperature dependence of magnetic $J_c$ is plotted. Modestly large transport $J_c$ at 20 K was also reported in the case of (Sr,Na)Fe$_2$As$_2$ tape [42]. Higher performance of (*AE*,Na)Fe$_2$As$_2$ around 20 K is more advantageous for the operation of them using He-free refrigeration systems in future. Third, it is possible that weak links between superconducting grains are improved by proper selections of a metal sheath and additives to the core, which is related to the reactivity of the compounds with Na. It was reported that presence of Ag-As compounds generated by the reaction of Ag sheath and As contained compounds in the core of (Sr,Na)Fe$_2$As$_2$ tape probably served to connect the grains and improved $J_c$ [41,42]. Enhancement of $J_c$ by the reaction between the sheath material and the superconducting compounds in the core have not been reported for (*AE*',K)Fe$_2$As$_2$ wires and tapes. Although the weak links have been improved by adding the materials such as Ag or Sn in (*AE*',K)Fe$_2$As$_2$, there have been no reports for (*AE*,Na)Fe$_2$As$_2$ wires and tapes. Solving these issues related to weak links based on the reaction between the core and the sheath material and additives, further enhancement of the $J_c$ can be expected. Finally, we discuss the possible enhancement of $J_c$ by the control of carrier concentration of Na and crystal structure in this system. It is reported that the magnetic $J_c$ of Ba$_{1-x}$K$_x$Fe$_2$As$_2$ single crystals shows a significant doping dependence, where the maximum value of $J_c$ appears at around $x \sim 0.3$ corresponding to slightly underdoped region compared with the optimal doping level of $x \sim 0.4$ [56, 57]. It was suggested that the origin of the characteristic doping dependence of $J_c$ is related to the orthorhombic-tetragonal phase boundary [58], which may work as pinning center to increase $J_c$. Some reports suggested the possible enhance of the $J_c$ in underdoped (Ba,K)Fe$_2$As$_2$ PIT wires [59,60]. On the other hand, Ba$_{1-x}$Na$_x$Fe$_2$As$_2$ also shows a similar orthorhombic-tetragonal phase transition [61]. This structural phase transition is reported to appear at $x \sim 0.36$, which is very close to the optimal doping level of $x \sim 0.4$ and $T_c$ is similar. Reduction of Na content of the starting powder of (Ba,Na)Fe$_2$As$_2$ of the PIT wires and tapes may effectively enhance the $J_c$.

In spite of their characteristic features and promising performance of (*AE*,Na)Fe$_2$As$_2$ as a raw material of the IBS wires and tapes, studies on (*AE*,Na)Fe$_2$As$_2$ wires and tapes have not been undertaken extensively compared with (*AE*',K)Fe$_2$As$_2$ wires and tapes, where the largest transport $J_c$ in the tape form has been reported. Only very recently, the fabrication (Ba,Na)Fe$_2$As$_2$ tapes has been reported for the first time [45]. Further enhancement of $J_c$ with additional purification of raw (*AE*,Na)Fe$_2$As$_2$ material and improvements of the condition of wire fabrication process are eagerly expected.

**Conclusion**

(Ba,Na)Fe$_2$As$_2$ superconducting round wires were fabricated by the PIT and HIP methods. By improving powder synthesis processes compared with the previous studies, highly pure raw materials for the wire fabrication were obtained. Superconducting properties for several wires were investigated. The largest transport critical current density ($J_c$) reached 40 kAcm$^{-2}$ at $T = 4.2$ K under a magnetic field of 100 kOe. This value exceeds not only the value of transport $J_c$ of previous (Ba,Na)Fe$_2$As$_2$ wire but also those of all iron-based superconducting round wires. Improvement of polycrystalline powder synthesis plays a key role for the enhancement of $J_c$ and wire fabrication processes have also affected their performance. Furthermore, it was clarified that stronger densification by high-pressure sintering and the texturing of grains in the wire core during the drawing also increase $J_c$ effectively. Differences and similarities between (*AE*,Na)Fe$_2$As$_2$ and more extensively studied (*AE*',K)Fe$_2$As$_2$ were discussed, and future prospects of the increase in $J_c$ in

($AE$,Na)Fe$_2$As$_2$ wires and tapes were suggested.

**Acknowledgements**

A part of this study was performed at the High Field Laboratory for Superconducting Materials, Institute for Materials Research, Tohoku University (Project No. 18H0011 and No. 19H0006). This work was partially supported by a Grant-in-Aid for Scientific Research (A) (17H01141) from the Japan Society for the Promotion of Science (JSPS).

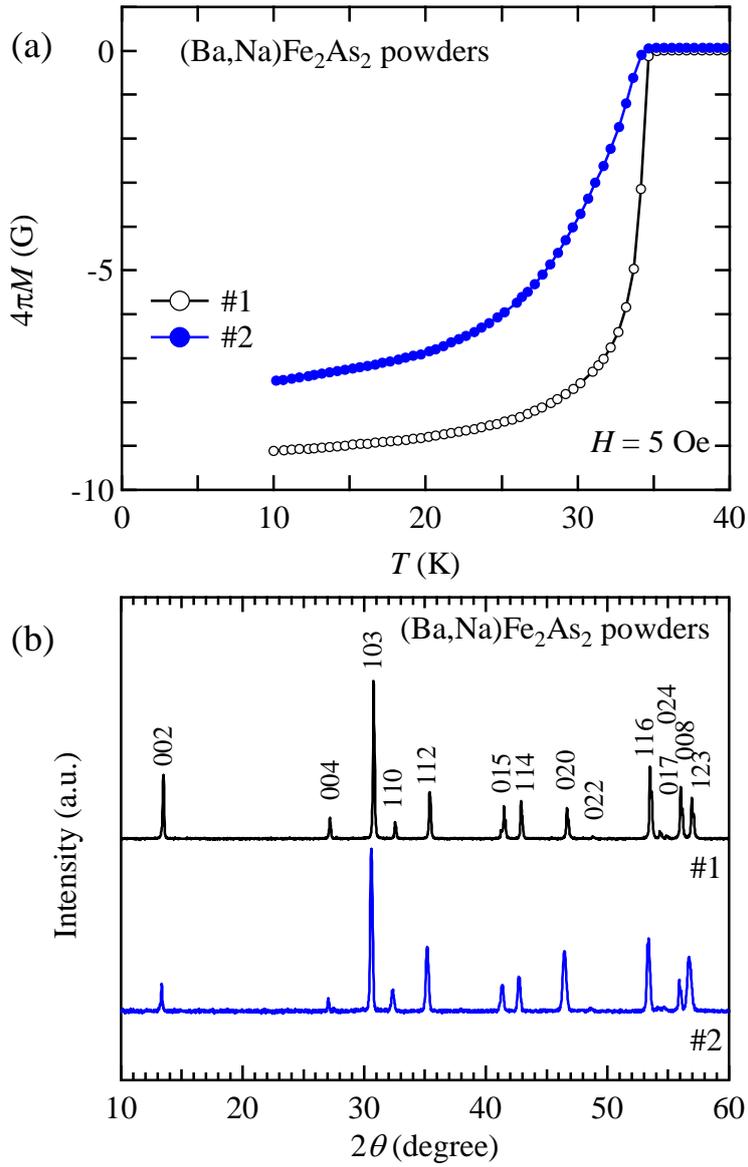

Figure 1. (a) Temperature dependence of magnetization at 5 Oe for (Ba,Na)Fe$_2$As$_2$ powder #1 and #2. (b) Powder XRD patterns of (Ba,Na)Fe$_2$As$_2$ powder #1 and #2.

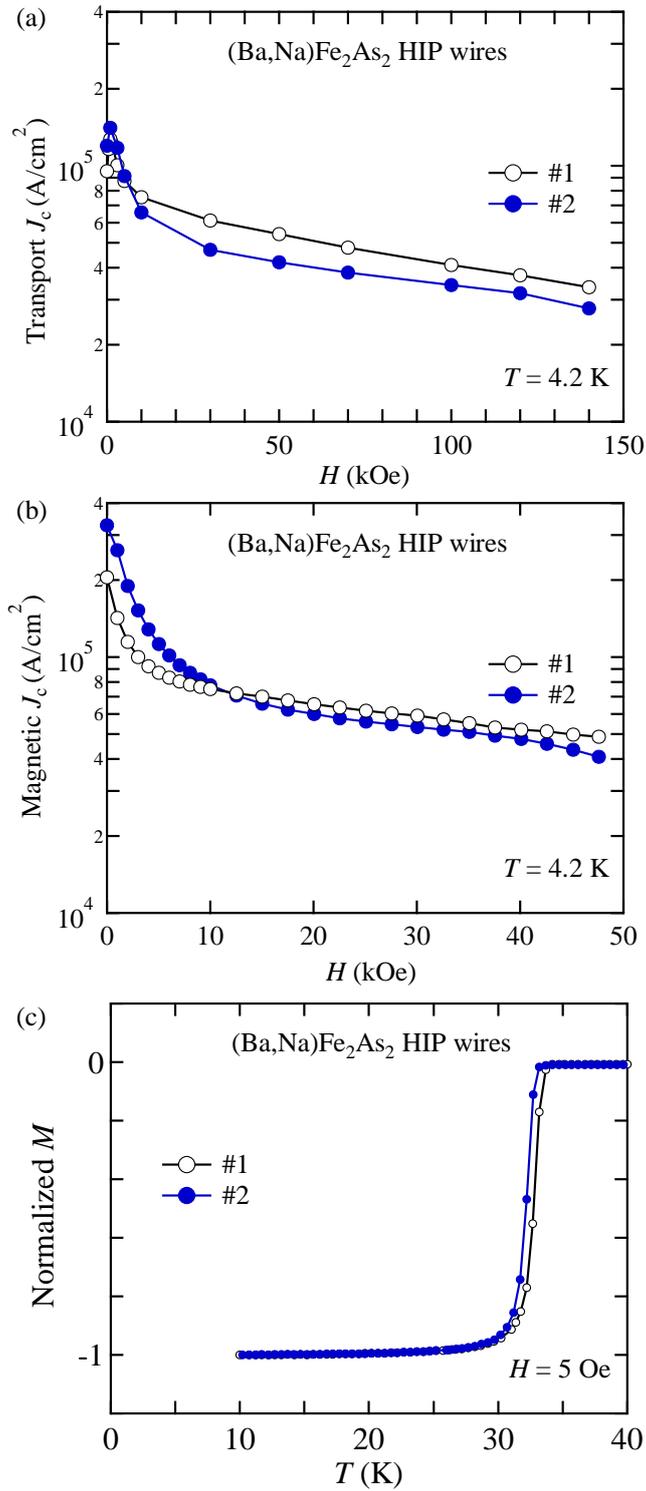

Figure 2. Magnetic field dependence of (a) the transport $J_c$ (*I-V*) and (b) magnetic $J_c$ (*M-H*) at 4.2 K for the $(Ba,Na)Fe_2As_2$ HIP wire #1 and #2. (c) Temperature dependence of normalized magnetization at 5 Oe for $(Ba,Na)Fe_2As_2$ HIP wire #1 and #2.

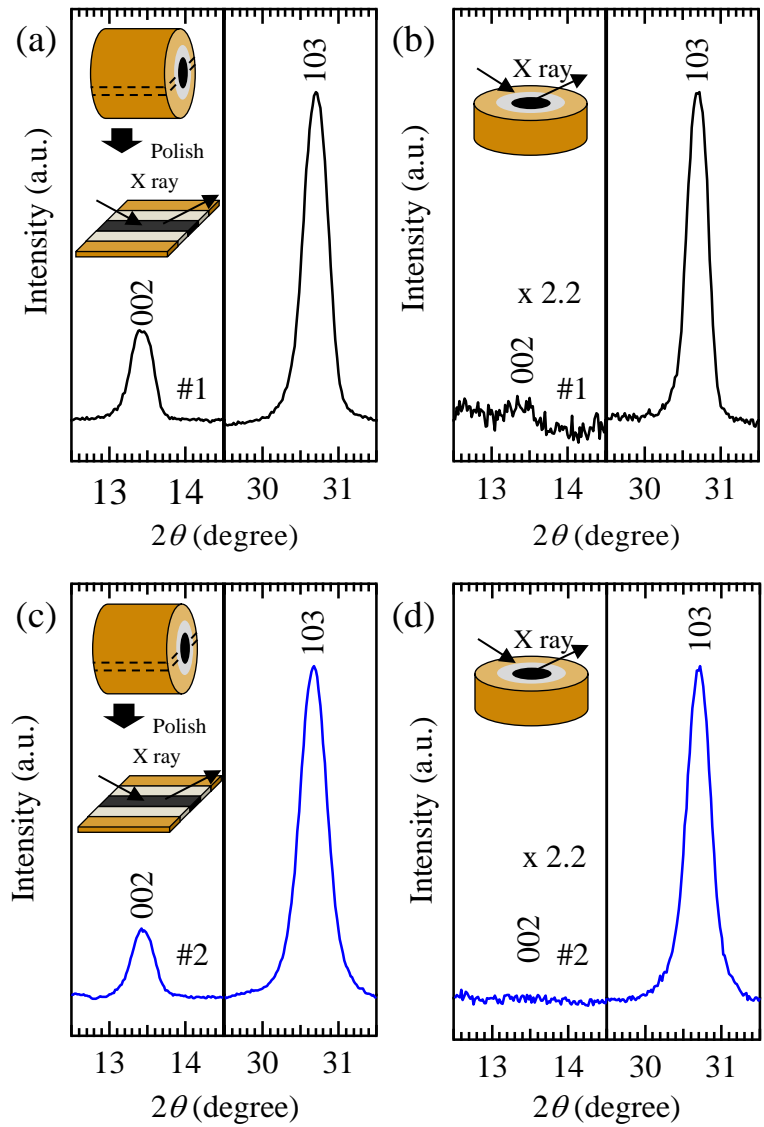

Figure 3. XRD patterns of (002) and (103) peaks for two different cross sections of polished HIP wire #1 ((a) and (b)) and #2 ((c) and (d)). Insets show the schematics of each cross section. Intensities in the left panel of (b) and (d) is multiplied by 2.2 to normalize the incident X-ray intensity for the comparison with the right panel of (b) and (d), respectively.

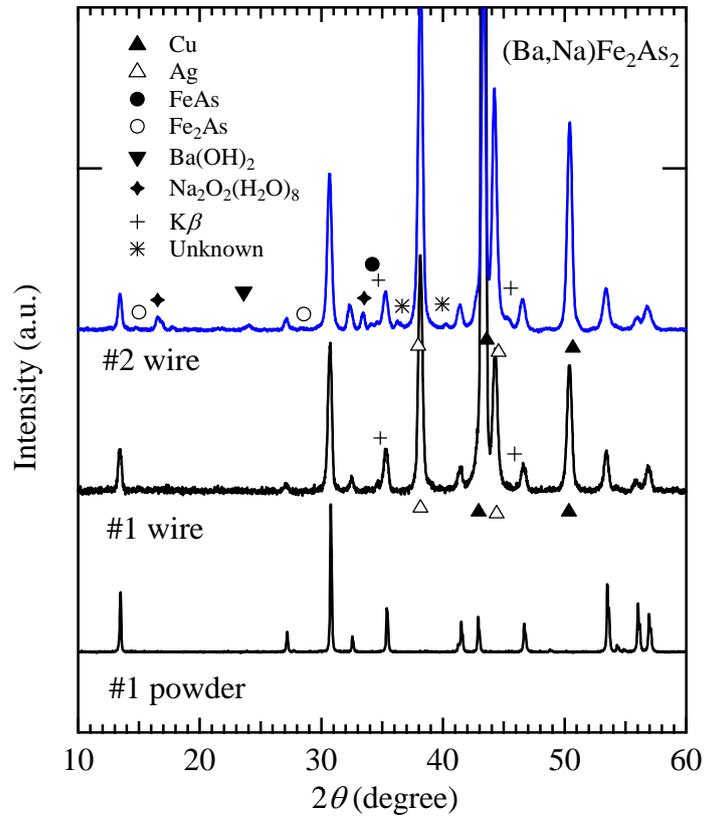

Figure 4. XRD patterns for the polished surface of the (Ba,Na)Fe$_2$As$_2$ HIP wire #1 and #2, and the (Ba,Na)Fe$_2$As$_2$ powder #1.

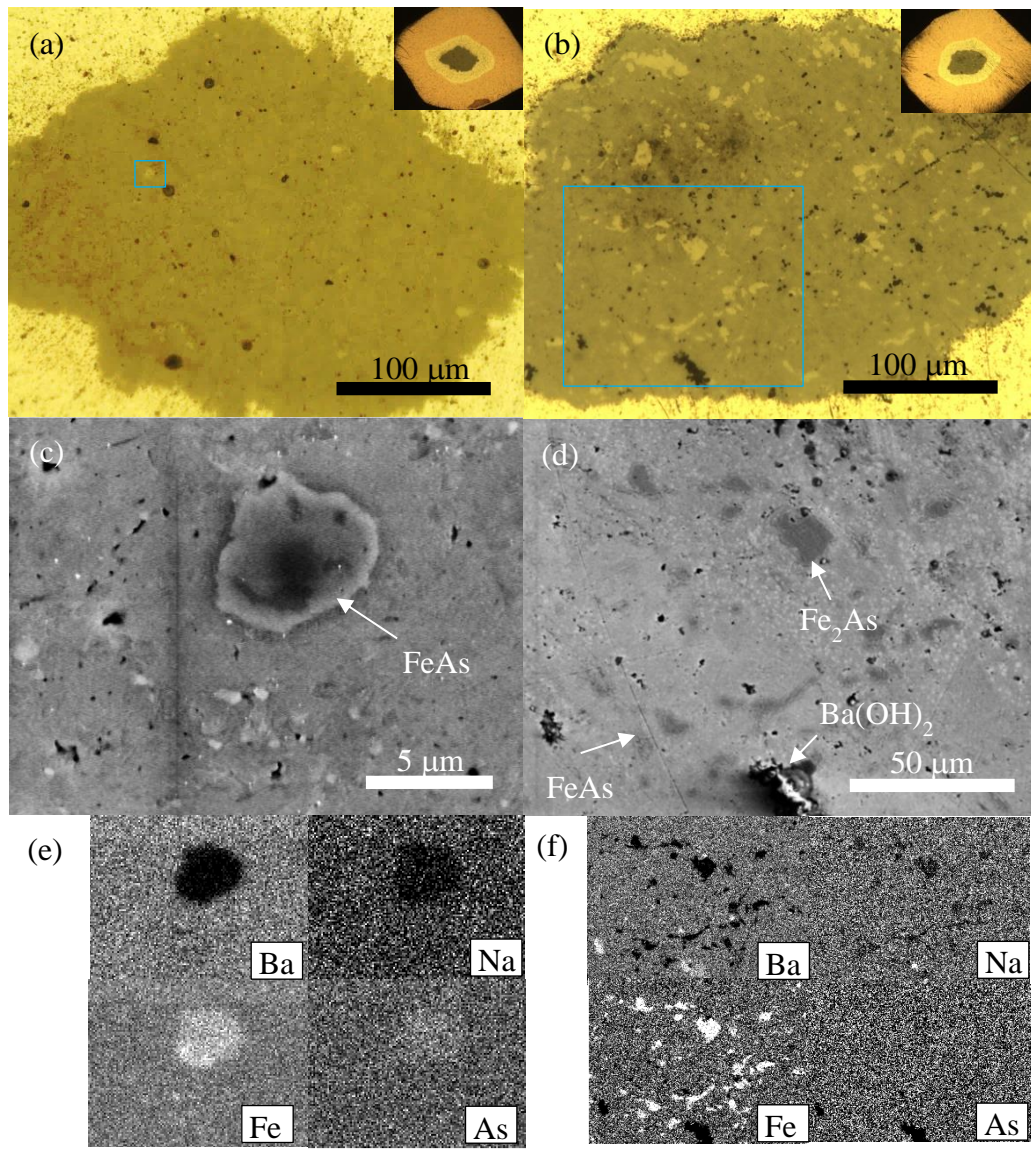

Figure 5. (a) Optical micrographs of the core of (Ba,Na)Fe$_2$As$_2$ HIP wire (a) #1 and (b) #2. Insets of (a) and (b) show the cross section of HIP wires. SEM images of the core of (Ba,Na)Fe$_2$As$_2$ HIP wire (c) #1 and (d) #2, which correspond to regions in blue rectangular boxes in (a) and (b), respectively. (e), (f) Elemental mappings for Ba, Na, Fe, and As, in the same region as (c) and (d), respectively.

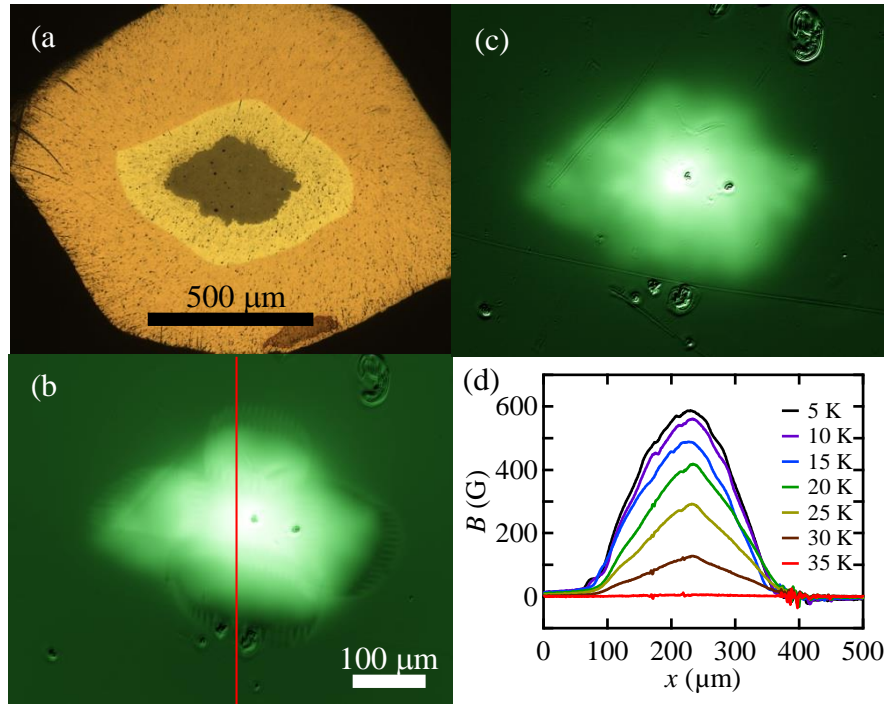

Figure 6. (a) An optical micrograph, MO images in the remanent state at (b) 5 K and (c) 30 K of (Ba,Na)Fe$_2$As$_2$ HIP wire #1. (d) Local magnetic induction profiles at different temperatures taken along the red line in (b).

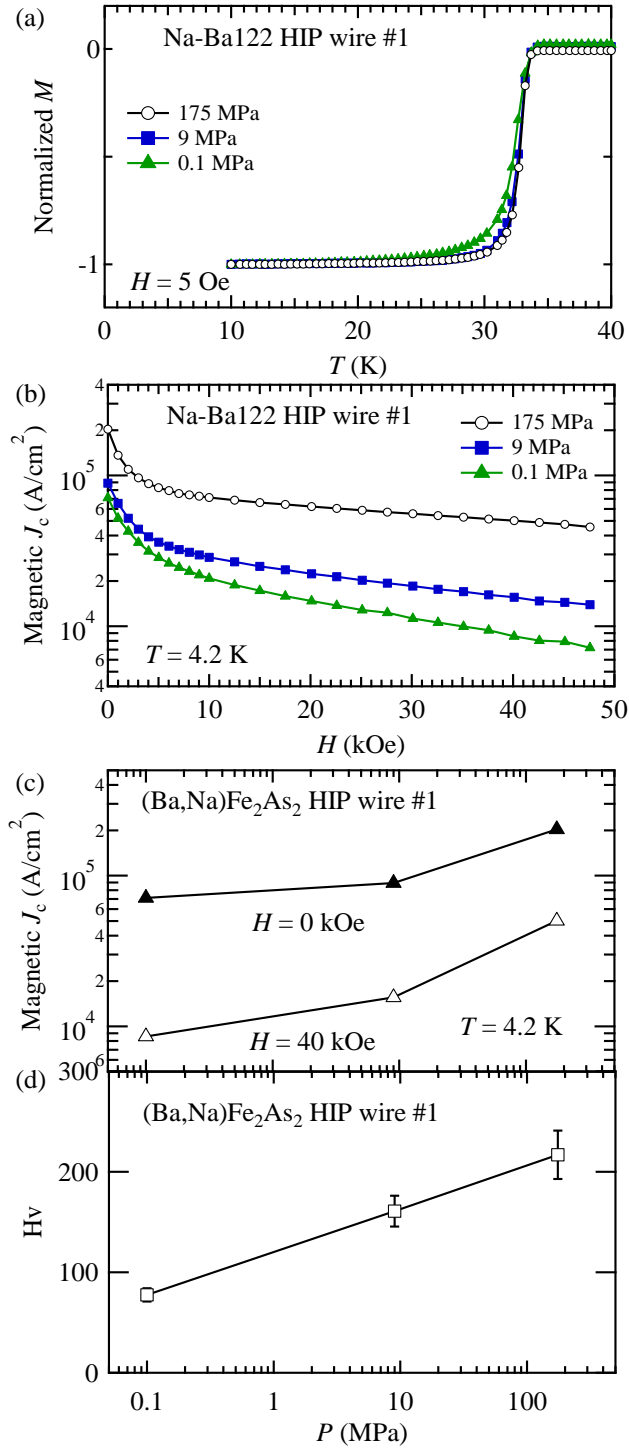

Figure 7. (a) Temperature dependence of normalized magnetization at 5 Oe and (b) magnetic field dependence of magnetic $J_c$ for $(Ba,Na)Fe_2As_2$ HIP wire #1, sintered at 700°C for 4 h at several pressures. Sintering pressure dependence of (c) magnetic $J_c$ at 4.2 K under self-field and at 40 kOe, and (d) Hv in the $(Ba,Na)Fe_2As_2$ HIP wire #1.

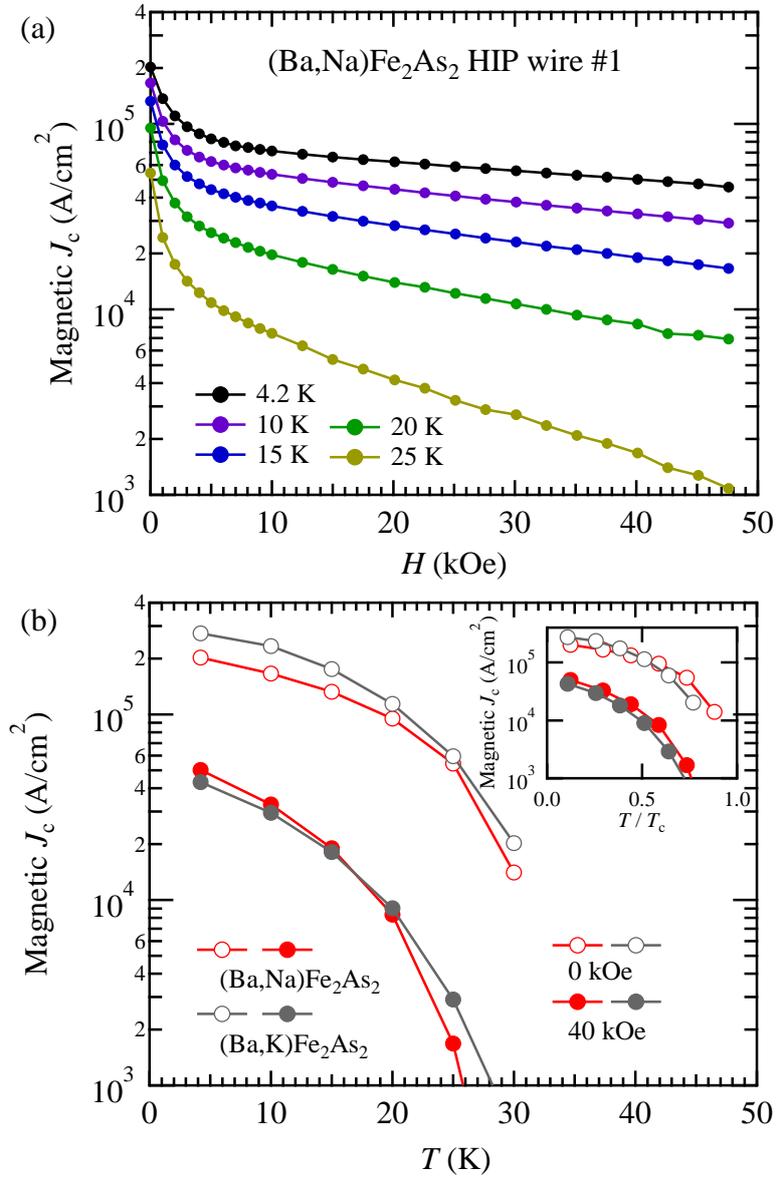

Figure 8. (a) Magnetic field dependence of magnetic $J_c$ between 4.2 K and 25 K of $(Ba,Na)Fe_2As_2$ HIP wire #1. (b) Temperature dependence of magnetic $J_c$ of $(Ba,Na)Fe_2As_2$ HIP wire #1 and $(Ba,K)Fe_2As_2$ HIP wire under self-field and at 40 kOe. The inset shows the normalized temperature dependence of them.